\documentclass[manuscript]{aastex}
\usepackage{spr-astr-addons}

\newcommand{\bq}{\begin{equation}}
\newcommand{\eq}{\end{equation}}
\newcommand{\bqn}{\begin{eqnarray}}
\newcommand{\eqn}{\end{eqnarray}}
\newcommand{\nb}{\nonumber}
\newcommand{\lb}{\label}
\newcommand{\rr}{\bf r}

\begin{document}
\title{Dynamical Evolution of an Unstable Gravastar with Zero Mass}
\shorttitle{Dynamical Evolution ...}
\shortauthors{Chan et al.}

\author{R. Chan \altaffilmark{1}}
\affil{Coordena\c{c}\~ao de Astronomia e Astrof\'{\i}sica, Observat\'orio
Nacional, Rua General Jos\'e Cristino, 77, S\~ao Crist\'ov\~ao, CEP 20921-400,
Rio
de Janeiro, RJ, Brazil}
\email{chan@on.br}
\and
\author{M.F.A. da Silva \altaffilmark{2}}
\affil{Departamento de F\'{\i}sica Te\'orica,
Instituto de F\'{\i}sica, Universidade do Estado do Rio de Janeiro,
Rua S\~ao Francisco Xavier 524, Maracan\~a,
CEP 20550-900, Rio de Janeiro - RJ, Brazil}
\email{mfasnic@gmail.com}
\author{Jaime F. Villas da Rocha \altaffilmark{3}} 
\affil{Universidade Federal do Estado do Rio de Janeiro,
Instituto de Bioci\^encias,
Departamento de Ci\^encias Naturais, Av. Pasteur 458, Urca,
CEP 22290-240, Rio de Janeiro, RJ, Brazil}
\email{jfvroch@pq.cnpq.br}
 
\date{\today}

\begin{abstract}
Using the conventional gravastar model, that is, an object constituted by two components 
where one of them is a massive infinitely thin shell and the other one is a de Sitter interior 
spacetime, we physically interpret a solution characterized by a
zero Schwarzschild mass. No stable gravastar is formed and it collapses without
forming an event horizon, originating what we call a massive non-gravitational
object.  The most surprise here is that the collapse occurs with an exterior de Sitter vacuum spacetime.
This creates an object which does not interact gravitationally
with an outside test particle and it may evolve to a point-like topological
defect.

\end{abstract}

\keywords{gravastar; black hole}

\section{Introduction}

Gravastars were proposed as an alternative model to black holes by Mazur and 
Motola \citep{Mazur02}. In their original form, they 
consisted of five layers: an internal core, $0 < R < R_1$ 
(where $R$ denotes the radius of the star), described by the de 
Sitter universe, an intermediate thin layer of stiff fluid, $R_1 < R < R_2$, 
an external region, $R > R_2$, described by the Schwarzschild solution, and two
infinitely thin shells, appearing, respectively, on the hypersurfaces $R = R_1$
and $R = R_2$ \citep{MM01}\citep{MM04}. The intermediate layer substituted the region 
where both horizons (of de Sitter and of Schwarzschild) should be present. 
Later, Visser and Wiltshire \citep{Visser04} simplified this model, reducing to 
three the number of regions. 
The first work to consider gravastar solutions
with de Sitter exterior were analyzed in \citep{Carter}.  In this work we will
use the same kind of exterior spacetime \citep{JCAP3}.
In recent works, two of us showed that 
gravastars do not substitute black holes \citep{JCAP}\citep{JCAP1}\citep{JCAP2}
\citep{JCAP3}\citep{JCAP4}. 
Instead of, they can coexist. In these studies were analyzed 
the stability of many gravastar's configurations. Now we are interested in a particular case,
 with zero Schwarzschild mass, which implies in a non-gravitational 
object. This structure is possible since the gravitational mass depends on 
the trace of the energy momentum tensor, instead of the energy density only. 
As the inner region is filled by dark energy, there is an equilibrium 
between the repulsive gravitational mass (the inner region) and the attractive 
one (the thin shell). 
This kind of objects has also been studied by two of us, in the case of a
charged shell \citep{JCAP4}.
In addition, we know that topological defects are 
extended solutions from Field Theory when the vacuum structure is topologically 
nontrivial. In three spatial dimensions 
one can have either point-like defects (monopoles), line-like defects (strings) 
or membrane-like defects (domain walls). 
So, the object that we study here is similar to a point-like topological 
defect, since the de Sitter vacuum solution does not describe 
all the spacetime. It is the aim of this work to investigate in details 
the properties of these interesting objects.

In this work we try to answer the 
following questions: what
kind of object would be formed and what would be the consequences of its existence, since
it collapses but without forming a black hole. After a detailed  analysis of
the physical interpretation, we concluded that a point-like zero mass
object, with non-gravitational interaction, is formed at the end of the process
and its existence is possible due the presence of an external cosmological constant.

The paper is organized as follows.  In Section II we present the gravastar 
model with an exterior de Sitter-Schwarzschild spacetime.
In Section III we justify the construction of a model of a non-gravitational
object using the calculus of the
total gravitational mass for a spherically symmetric system, applied to the
gravastar solution considered.
In Section IV we obtain the potential of the 
gravastar with $m=0$ (that can evolve to a point-like topological defect), that 
describes the dynamical evolution of these objects. Finally, in Section V we 
present the final remarks.

\section{Gravastar Model with $m=0$}

In this section, we will summarize the gravastar model with 
an exterior de Sitter-Schwarzschild spacetime \citep{JCAP3}.
Since we assume that total energy of the system is zero,  
the motion of the surface
of the gravastar can be written in the form \citep{Visser04},
\bq
\lb{1.4}
\frac{1}{2}\dot{R}^{2} + V(R) = 0,
\eq 
where $V(R)$ is the potential and $\dot{R} \equiv dR/d\tau$, with
$\tau$ being the proper time of the surface. 

This potential presents two particularly interesting types of profiles. 
A group of them is presented in the figure \ref{fig1}-a; 
where from top 
to bottom, depending on the initial radius of the shell, the first solid curve 
represents a stable gravastar configuration, the second solid curve shows the 
bounded excursion stable gravastar \citep{Visser04}\citep{JCAP}. In the figure 
\ref{fig1}-b, 
we show the profiles which represent collapsing configurations; depending also
on the initial radius, we can have dispersion of the shell (top dashed line), 
collapse of the shell (bottom dashed line) and the critical profile,  transition
between the two behaviors (solid line).

\begin{figure}
\epsscale{.80}
\plotone{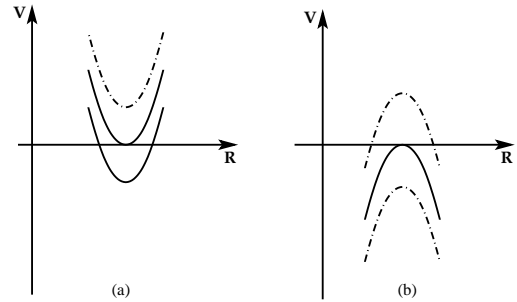}
\caption{The possible potentials $V(R)$.}
\lb{fig1}
\end{figure}

The interior spacetime is described by the de Sitter metric, given by 
\bq
ds^2_{i}=-f_1 dt^2 + f_2 dr^2 + r^2 d\Omega^2,
\lb{ds2-}
\eq
where $f_1=1- (r/L_i)^2$,
$f_2= \frac{1}{1 - (r/L_i)^2}$, $L_i=\sqrt{3/\Lambda_i}$ and 
$d\Omega^2 = d\theta^2 + \sin^2(\theta)d\phi^2$. Hereinafter,
this spacetime with the cosmological constant $\Lambda_i$ will be interpreted 
as a fluid due to dark energy.

The exterior spacetime is given by a de Sitter-Schwarzschild metric with $m=0$, that is
\bq
ds^2_{e}= - f dv^2 + f^{-1} d{\rr}^2 + {\rr}^2 d\Omega^2,
\lb{ds2+}
\eq
where $f=1 -({\rr}/L_e)^2$ and $L_e=\sqrt{3/\Lambda_e}$.

The metric of the hypersurface of the shell is given by
\bq
ds^2_{\Sigma}= -d\tau^2 + R^2(\tau) d\Omega^2.
\lb{ds2Sigma}
\eq

Since $ds^2_{i} = ds^2_{e} = ds^2_{\Sigma}$ then $r_{\Sigma}={\rr}_{\Sigma}=R$,
and besides
\bqn
\dot t^2&=& \left[ 1- \left(\frac{R}{L_i}\right)^2+ \dot R^2 \right] \left[1- 
\left(\frac{R}{L_i}\right)^2 \right]^{-2},
\lb{dott2}
\eqn
and
\bqn
\dot v^2&=& \left[ 1-\left(\frac{R}{L_e}\right)^2+ \dot R^2 \right] 
\left[1-\left(\frac{R}{L_e}\right)^2 \right]^{-2},
\lb{dotv2}
\eqn
where the dot represents the differentiation with respect to $\tau$.

Thus, the interior and exterior normal vectors are given by
\bq
n^{i}_{\alpha} = (-\dot R, \dot t, 0, 0 ),\qquad n^{e}_{\alpha} = 
(-\dot R, \dot v, 0, 0 ).
\lb{nalpha-}
\eq

In addition, the interior and exterior extrinsic curvatures are given by
\bqn
K^{i}_{\tau\tau}&=&-[(3 L_i^4 \dot R^2-L_i^4 \dot t^2+2 L_i^2 R^2 \dot t^2-
R^4 \dot t^2) R \dot t- \nb \\
& &(L_i+R) (L_i-R) (\dot R \ddot t-\ddot R \dot t) L_i^4] \times \nb \\
& &(L_i+R)^{-1} (L_i-R)^{-1} L_i^{-4}
\lb{Ktautau-}
\eqn
\bq
K^{i}_{\theta\theta}=\dot t (L_i+R) (L_i-R) L_i^{-2} R
\lb{Kthetatheta-}
\eq
\bq
K^{i}_{\phi\phi}=K^{i}_{\theta\theta}\sin^2(\theta),
\lb{Kphiphi-}
\eq
\bqn
K^{e}_{\tau\tau}&=&\dot v [(L_e^2 R \dot R-L_e^2 R \dot v+R^
3 \dot v) (-L_e^2 R \dot R-L_e^2 R \dot v+ \nb \\
& &R^3 \dot v)- \nb \\
& &2 L_e^4 R^2 \dot R^2] ((-R) L_e^2+R^ 3)^{-1}
(-R^3) L_e^{-4} R^{-3}+ \nb \\
& &\dot R \ddot v- \ddot R \dot v
\lb{Ktautau+}
\eqn
\bq
K^{e}_{\theta\theta}= -\dot v((-R) L_e^2+R^3) L_e^{-2}
\lb{Kthetatheta+}
\eq
\bq
K^{e}_{\phi\phi}=K^{e}_{\theta\theta}\sin^2(\theta).
\lb{Kphiphi+}
\eq

Considering the  interior and exterior extrinsic curvature, 
$K_{\alpha\beta} $, we have \citep{Lake}
\bq
[K_{\theta\theta}]= K^{e}_{\theta\theta}-K^{i}_{\theta\theta} = - M,
\lb{M}
\eq
where $M$ is the mass of the shell. Thus, it can be written as
\bq
\lb{M1}
M = -\dot v R\left[ 1-\left(\frac{R}{L_e}\right)^2 \right]+
\dot t R \left[1- \left(\frac{R}{L_i}\right)^2 \right] . 
\eq

Then, substituting equations (\ref{dott2}) and (\ref{dotv2}) into (\ref{M1})
we get
\bqn
M&=&R \left[ 1 -\left(\frac{R}{L_i}\right)^2  + \dot R^2 \right]^{1/2}- \nb \\
& &R\left[1 -\left(\frac{R}{L_e}\right)^2 + \dot R^2 \right]^{1/2}.
\lb{M2}
\eqn

In order to keep the ideas of our work \citep{JCAP1} as much as possible, we 
consider the thin shell as consisting
of a fluid with an equation of state, $\vartheta = (1-\gamma)\sigma$, where 
$\sigma$ and $\vartheta$ denote, 
respectively, the surface energy density and pressure of the shell and $\gamma$
is a constant. 
The equation of motion of the shell can be written as \citep{Lake}
\bqn
\dot M + 8\pi R \dot R \vartheta &=& 4 \pi R^2 [T_{\alpha\beta}u^{\alpha}n^{\beta}]= \nb \\
& &4 \pi R^2 \left(T^e_{\alpha\beta}u_e^{\alpha}n_e^{\beta}-T^i_{\alpha\beta}
u_i^{\alpha}n_i^{\beta} \right),
\lb{dotM}
\eqn
where $u^{\alpha}$ is the four-velocity.  Since the interior 
and the exterior spacetimes correspond to vacuum solutions, we get
\bq
\dot M + 8\pi R \dot R (1-\gamma)\sigma = 0,
\lb{dotM1}
\eq
and since $\sigma = M/(4\pi R^2)$ we can solve equation (\ref{M1}) giving
\bq
M=k R^{2(\gamma-1)},
\lb{Mk}
\eq
where $k$ is an integration constant.

Substituting equation (\ref{Mk}), into 
(\ref{M2}) and solving it for $\dot R^2/2$ we obtain the potential 
$V(R,L_i,L_e)$, given by \citep{JCAP3}
\bqn
& &V(R,L_i,L_e,\gamma)= -\frac{1}{2}\left[-1 +\frac{R^{(4\gamma-6)}}{4}+
\frac{R^2}{2L_i^2}+ \right. \nb \\
& &\left. \frac{R^{(-4\gamma+10)}}{4L_i^4}- \frac{R^{(-4\gamma+10)}}{2L_i^2L_e^2}+ 
\frac{R^2}{2L_e^2}+ \right. \nb \\
& &\left. \frac{R^{(-4\gamma+10)}}{4L_e^4} \right],
\lb{VR}
\eqn
where we have redefined the cosmological constants 
$L_i$ and $L_e$ and the radius $R$ as  
$L_i \equiv L_i k^{\frac{2}{2\gamma-3}} $, 
$L_e \equiv L_e k^{\frac{2}{2\gamma-3}}$, 
$R \equiv Rk^{-\frac{1}{2\gamma-3}} $.

Therefore, for any given constants $L_e$, $L_i$ and equation (\ref{VR}) uniquely
determines the collapse of the prototype gravastar.
Depending on the initial value $R_0$, the shell can stay stable and 
form  a gravastar, the shell can collapses to a finite non-zero minimal radius and 
then expands to infinity, transforming the whole spacetime into a de Sitter spacetime, or
collapses in a structure without horizon.
To  guarantee
that initially the spacetime does not have any kind of horizons, cosmological 
or event, we must restrict $R_{0}$ to the ranges simultaneously,
\bq
\lb{2.2b}
0<R_{0} < L_i,\qquad 0<R_{0} < L_e,
\eq
where $R_0$ is the initial collapse radius.

\section{Total Gravitational Mass}

In order to study the gravitational effect generated by the two components of the
gravastar, i.e., the interior de Sitter and the thin shell in the exterior region,
we need to calculate the total gravitational mass of a spherical symmetric system. 
Some alternative definitions are given by \citep{Marder},\citep{Israel} and \citep{Levi}. 
Here we consider the Tolman's formula for the mass, which is given by 
\bq
M_{G}=\int_0^{R_0} \int_{-\pi}^{\pi} \int_0^{2\pi} \sqrt{-g}\; 
T^\alpha_\alpha dr d\theta d\phi,
\eq
where $\sqrt{-g}$ is the determinant of the metric.  For the special case of a 
thin shell we have
\bq
M_{G}=\int_0^{R_0} \int_{-\pi}^{\pi} \int_0^{2\pi} \sqrt{-g}\; 
T^\alpha_\alpha \delta({\rr}-R_0) d{\rr} d\theta d\phi.
\eq

Thus, the Tolman's gravitational mass of the thin shell is given by
\bq
M_G^{shell}=(3-2\gamma)M,
\eq
and for the interior de Sitter (dS) spacetime we have
\bq
M_{G}^{dS}=-\frac{2}{3} \Lambda_i R_0^3.
\eq

Note that for $0 \le \gamma \le 3/2$ we have standard energy and 
$M_G^{shell} \ge 0$, and for
$\gamma > 3/2$ we get dark energy and $M_G^{shell} < 0$. On the other hand, 
the de Sitter
interior presents a negative gravitational mass, since $\Lambda_i > 0$, 
in agreement with its repulsive effect.

Now we can write the total Tolman's gravitational mass of the gravastar as
\bq
\label{MGtotal}
M_G^{total}=M_G^{shell}+M_G^{dS}= (3-2\gamma)M-\frac{2}{3}\Lambda_i R_0^3.
\eq

This mass also should represent the Schwarzschild exterior mass ($m=M_G^{total}$) 
for this gravastar. Note that for the parameters used in the configuration 
described by the figure 7, we have that $M$(the shell's mass)$=8/9$.
Based on this, it is possible to obtain a physical structure through a 
combination of these two solutions,
which results in a system with $m=0$, a de Sitter-Schwarzschild exterior 
spacetime. 
Then, an object like this collapses without forming a black hole. Instead 
of it generates an exterior spacetime which is locally a de Sitter 
spacetime.  A test particle would not interact gravitationally with the 
central source, similar to what happens with the cosmic strings. A point-like topological 
defect seems to be present here.

\section{The Analysis of the Potential}

For the particular case where the Schwarzschild mass is vanished, $m=0$, 
the equation (\ref{M2}) 
allows us to see clearly that a positive mass of the shell $M$ is possible only 
if  $L_e < L_i$ (or $\Lambda_i < \Lambda_e$).
As a consequence, objects similar to these ones are allowed only with an exterior
de Sitter vacuum spacetime, but never with a Minkowski spacetime.
The mass of the shell also revels that, in the limit $R\rightarrow 0$, it is
infinitesimally small, but different from zero, since $R$ cannot be zero,
in order to preserve the configuration of the object which has an interior
de Sitter spacetime. Then, from the collapse, necessarily an object is formed
with a de Sitter interior spacetime surrounded by a massive thin shell,
which can be matched with a de Sitter exterior spacetime.
Note also that if $L_e = L_i$, the mass of the shell is zero and no object is formed.
Recently we have shown that in order to have a physically 
acceptable stable shell between two spacetimes with cosmological constants, the inner 
cosmological constant must be greater than that outer one \citep{JCAP3}, which is in 
agreement to the gravastar requirement proposed by Horvat \& Ilijic \citep{grava1}. 
Our present work is completely coherent with this previous result since here it 
is not possible to have a stable thin shell.

Since $L_e < L_i$ then we can define $\alpha = L_e/L_i$ ($0 < \alpha < 1$) and we can rewrite equation (\ref{VR}) as
\bqn
& &V(R,L_i,\alpha,\gamma)= \frac{1}{2}-\frac{1}{4}\frac{R^2}{L_i^2}-\frac{1}{8}\frac{R^{10}}{L_i^4\alpha^4 R^{4\gamma}}+\nb \\
& &\frac{1}{4}\frac{R^{10}}{L_i^4\alpha^2 R^{4\gamma}}-\frac{1}{4}\frac{R^2}{L_i^2\alpha^2}-\frac{1}{8}\frac{R^{10}}{L_i^4R^{4\gamma}}-\frac{1}{8}R^{4\gamma-6}.
\lb{V2}
\eqn

A first differentiation of the potential gives 
\bqn
& &\frac{dV}{dR}(R,L_i,\alpha,\gamma) = \nb \\
&=& \frac{3}{4}R^{4\gamma-7}-\frac{5}{4}\frac{R^9}{L_i^4 \alpha^4 R^{4\gamma}}-\frac{5}{4}\frac{R^9}{L_i^4R^{4\gamma}}+\frac{5}{2}\frac{R^9}{L_i^4\alpha^2R^{4\gamma}}+ \nb \\
& &\frac{1}{2}\frac{\gamma R^9}{L_i^4\alpha^4R^{4\gamma}}-\frac{\gamma R^9}{L_i^4\alpha^2R^{4\gamma}}- \nb \\
& & \frac{1}{2}\gamma R^{4\gamma-7}-\frac{1}{2}\frac{R}{L_i^2\alpha^2}- \frac{1}{2}\frac{R}{L_i^2}+\frac{1}{2}\frac{\gamma R^9}{L_i^4R^{4\gamma}}.
\lb{V3}
\eqn

A second differentiation of the potential furnishes
\bqn
& &\frac{d^2V}{dR^2}(R,L_i,\alpha,\gamma) = \nb \\
& &-2 \frac{\gamma^2 R^8}{ L_i^4 \alpha^4 R^{4\gamma}}-2 \frac{\gamma^2 R^{4\gamma}}{ R^8}-2 \frac{\gamma^2 R^8}{ L_i^4 R^{4\gamma}}-\frac{1}{2L_i^2\alpha^2}- \nb \\
& &\frac{1}{2 L_i^2}-\frac{45}{4} \frac{R^8}{L_i^4 \alpha^4 R^{4\gamma}}-\frac{45}{4} \frac{R^8}{L_i^4 R^{4\gamma}}- \nb \\
& &19 \frac{\gamma R^8}{L_i^4 \alpha^2 R^{4\gamma}}+\frac{45}{2} \frac{R^8}{L_i^4 \alpha^2 R^{4\gamma}}-\frac{21}{4} R^{4\gamma-8}+ \nb \\
& &\frac{13}{2} \gamma R^{4\gamma-8}+4 \frac{\gamma^2 R^8}{L_i^4 \alpha^2 R^{4\gamma}}+\frac{19}{2} \frac{\gamma R^8}{L_i^4 \alpha^4 R^{4\gamma}}+ \nb \\
& &\frac{19}{2} \frac{\gamma R^8}{L_i^4 R^{4\gamma}}.
\lb{V4}
\eqn

Vanishing of the first derivative of the potential gives us the critical 
values of the parameter $L_i$, or the values of $L_i$ which assure that the 
potential has a profile as showed in the figure \ref{fig1}. 
Doing this we obtain two solutions for $L_{ic}^2$
\bq
L_{ic1}^2= -R^{8 -4 \gamma} \left[ \frac{\alpha^2+1-2  \sqrt{ \Delta}}{(2 \gamma-3) \alpha^2} \right],
\lb{Lic21}
\eq
and
\bq
L_{ic2}^2= -R^{8 -4 \gamma} \left[ \frac{\alpha^2+1+2  \sqrt{ \Delta}}{(2 \gamma-3) \alpha^2} \right],
\lb{Lic22}
\eq
where $\Delta= 4 \alpha^4-7 \alpha^2+4+\gamma^2-4 \gamma-4 \gamma \alpha^4+8 \gamma \alpha^2-2 \gamma^2 \alpha^2+\gamma^2 \alpha^4$.
We can see from equations (\ref{Lic21}) and (\ref{Lic22}) that $L_{ic1}^2$ may be positive
or negative and $L_{ic2}^2$ is always positive, since for $-3/2 < \gamma < 3/2$ we have 
a standard energy shell. 
The next figures 2 to 5 are shown in order to determine the sign of some
functions. This could not be done analytically because the complex dependence
of these functions on $\gamma$ and $\alpha$.
In figure \ref{fig2} we can see that $\Delta$ is
always positive, showing that the solutions for $L_ic$ are always real.
The figures \ref{fig3} and \ref{fig4} show that the unique valid solution is $L_{ic}=L_{ic2}$.

\begin{figure}
\epsscale{.80}
\plotone{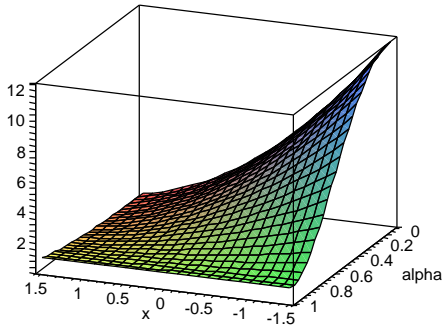}
\caption{This plot shows, in terms of $\alpha$ and $X \equiv \gamma$,
the function $\Delta$, in the intervals
$-3/2 <\gamma < 3/2$ and $0 < \alpha < 1$. This function is always positive
anywhere, this shows that the solutions for $L_{ic}$ are always real.}
\lb{fig2}
\end{figure}

\begin{figure}
\epsscale{.80}
\plotone{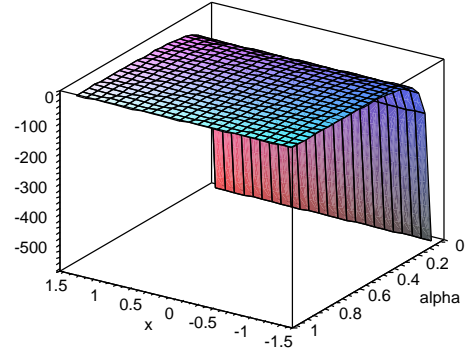}
\caption{This plot shows, in terms of $\alpha$ and $X \equiv \gamma$,
the function $L_{ic1}^2 R^{4\gamma-8}$, in the intervals
$-3/2 <\gamma < 3/2$ and $0 < \alpha < 1$. This function is always negative
anywhere, showing that this solution for $L_{ic}$ is not real.}
\lb{fig3}
\end{figure}

\begin{figure}
\epsscale{.80}
\plotone{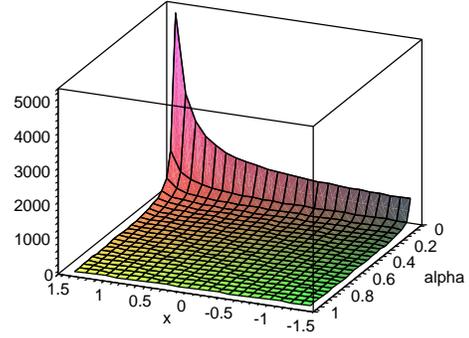}
\caption{This plot shows, in terms of $\alpha$ and $X \equiv \gamma$,
the function $L_{ic2}^2 R^{4\gamma-8}$, in the intervals
$-3/2 <\gamma < 3/2$ and $0 < \alpha < 1$. This function is always positive
anywhere, showing that this is the solution for $L_{ic}$.}
\lb{fig4}
\end{figure}

\begin{figure}
\epsscale{.80}
\plotone{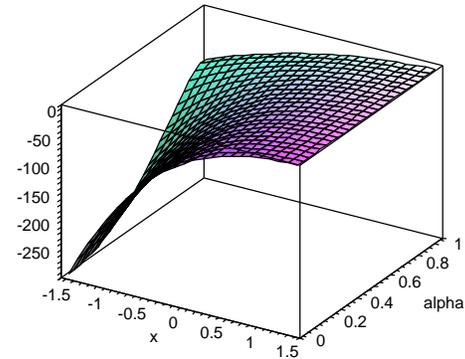}
\caption{This plot shows, in terms of $\alpha$ and $X \equiv \gamma$,
the second derivative of the potential $d^2V/dR^2(R,L_i=L_{ic},\alpha,\gamma) R^
{-4\gamma+8}$, in the intervals
$-3/2 <\gamma < 3/2$ and $0 < \alpha < 1$. Since the derivative is always negative
anywhere. This shows that the potential $V(R,L_i,\alpha,\gamma)$ has always maxi
ma.}
\lb{fig5}
\end{figure}

Substituting $L_i$ by $L_{ic}$ into the second derivative of the potential, 
equation (\ref{V4}), we get
\bqn
& &\frac{d^2V}{dR^2}(R,L_i=L_{ic},\alpha,\gamma) = \nb \\
& &- \frac{4R^{4\gamma-8}}{\alpha^2+1+2 \sqrt{\Delta}} \left( 48-7 \gamma \sqrt{\Delta}-164 \gamma^2 \alpha^2+ \right. \nb \\
& &\left. 84 \gamma^2 \alpha^4+84 \gamma^2+2 \gamma^2 \alpha^2 \sqrt{\Delta}+48 \alpha^4-84 \alpha^2+ \right. \nb \\
& & \left. 6 \alpha^2 \sqrt{\Delta}+2 \gamma^2 \sqrt{\Delta}-7 \gamma \alpha^2 \sqrt{\Delta}-30 \gamma^3+4 \gamma^4- \right. \nb \\
& &\left. 30 \gamma^3 \alpha^4-8 \gamma^4 \alpha^2+60 \gamma^3 \alpha^2+4 \gamma^4 \alpha^4+ \right. \nb \\
& & \left. 194 \gamma \alpha^2-104 \gamma \alpha^4-104 \gamma+6 \sqrt{\Delta} \right)
\eqn
The plot is shown in the figure \ref{fig5}. Then, we can see that,
the second derivative of the potential, in the intervals
$-3/2 <\gamma < 3/2$ and $0 < \alpha < 1$, is always negative
anywhere. This shows that the potential $V(R,L_i,\alpha,\gamma)$ has always maxima 
and the potential is well represented by the profiles of the figure \ref{fig1}-b.

Let us now, calculate the potential at $L_i=L_{ic}$ in order to determine
if it is positive or negative. Thus, we will consider three different
cases: a rigid fluid shell ($\gamma=0$), a dust shell ($\gamma=1$) and
another case of standard energy shell ($\gamma=5/4$).
Thus,
\bqn
& & V(R,L_i=L_{ic},\alpha,\gamma=0)= \nb \\
& & \frac{1}{2R^6 (\alpha^2+1+2 \sqrt{\Delta_1})^2} \left( -26 R^6 \alpha^2+17 R^6- \right. \nb \\
& &\left. 4 \alpha^2 \sqrt{\Delta_1}-8+4 R^6 \alpha^2 \sqrt{ \Delta_1}+8 \alpha^2-8 \alpha^4- \right. \nb \\
& & \left. 4 \sqrt{\Delta_1}+17 R^6 \alpha^4+4 R^6 \sqrt{\Delta_1} \right),
\eqn
where $\Delta_1=4 \alpha^4-7 \alpha^2+4$,
\bqn
& & V(R,L_i=L_{ic},\alpha,\gamma=1)= \nb \\
& & \frac{1}{2R^2 (\alpha^2+1+2 \sqrt{\Delta_2})^2} \left(-2 R^2 \alpha^2+ 5 R^2- \right. \nb \\
& &\left. 2 \alpha^2 \sqrt{\Delta_2}+4 R^2 \alpha^2 \sqrt{\Delta_2}-  2 \alpha^4-2 \sqrt{\Delta_2}+ \right. \nb \\
& & \left. 5 R^2 \alpha^4-2 +4 R^2 \sqrt{\Delta_2} \right)
\eqn
where $\Delta_2=\alpha^4-\alpha^2+1$,
\bqn
& & V(R,L_i=L_{ic},\alpha,\gamma=5/4)= \nb \\
& & \frac{1}{4R(2 \alpha^2 +2 +\sqrt{\Delta_3})^2} \left( 12 R \alpha^2+26 R-3 \alpha^2 \sqrt{\Delta_3}+ \right. \nb \\
& &\left. 8 R \alpha^2 \sqrt{\Delta_3}-6 \alpha^2 - 9 \alpha^4-3 \sqrt{\Delta_3}+ \right.  \nb \\
& & \left. 26 R \alpha^4-9+8 R \sqrt{\Delta_3} \right)
\eqn
where $\Delta_3=9 \alpha^4-2 \alpha^2+9$.

In the next figures 6 to 8 show the potential for three cases with different
equations of state ($\gamma=0,1,5/4$) in order to show that the gravastar model
treated in this work evolves dynamically to a collapse.

In figures \ref{fig6}, \ref{fig7} and \ref{fig8} we show the potential
calculated at $L_i=L_{ic}$ and $\gamma=0, 1, 5/4$, respectively.
Since the potential is always negative anywhere, this shows that the 
potential $V(R,L_i=L_{ic},\alpha,\gamma)$ represents a gravastar collapse.

The examples of these types of potential are shown in figure \ref{pot} for
three different values of $\alpha=0.1,$ $0.06470588235,$ $0.04117647059$.

Thus, solving equation (\ref{M2}) we can obtain $\dot R(\tau)$ and $R(\tau)$ which are
shown in the figure \ref{Rtau} for $\gamma=0$, $\alpha=0.1$ and $L_i=0.85$.

\begin{figure}
\epsscale{.80}
\plotone{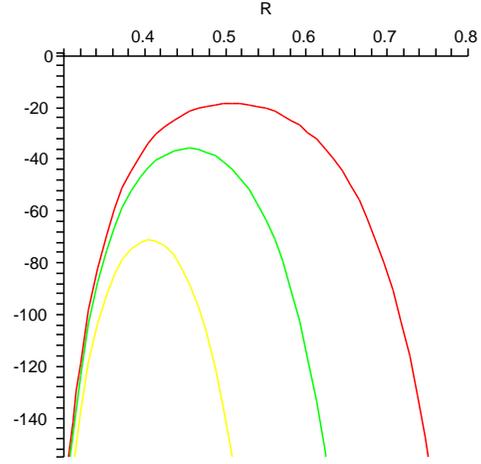}
\caption{This plot shows, in terms of $R$,
the potential $V(R,L_i=0.85,\alpha,\gamma=0)$, for
$\alpha=0.1,\; 0.06470588235,\; 0.04117647059$, represented by the top to bottom curves,
respectively.  Since the potential is always negative
anywhere, this shows that the potential $V(R,L_i=0.85,\alpha,\gamma=0)$ represents a
collapse.}
\lb{pot}
\end{figure}

\begin{figure}
\epsscale{1.60}
\plottwo{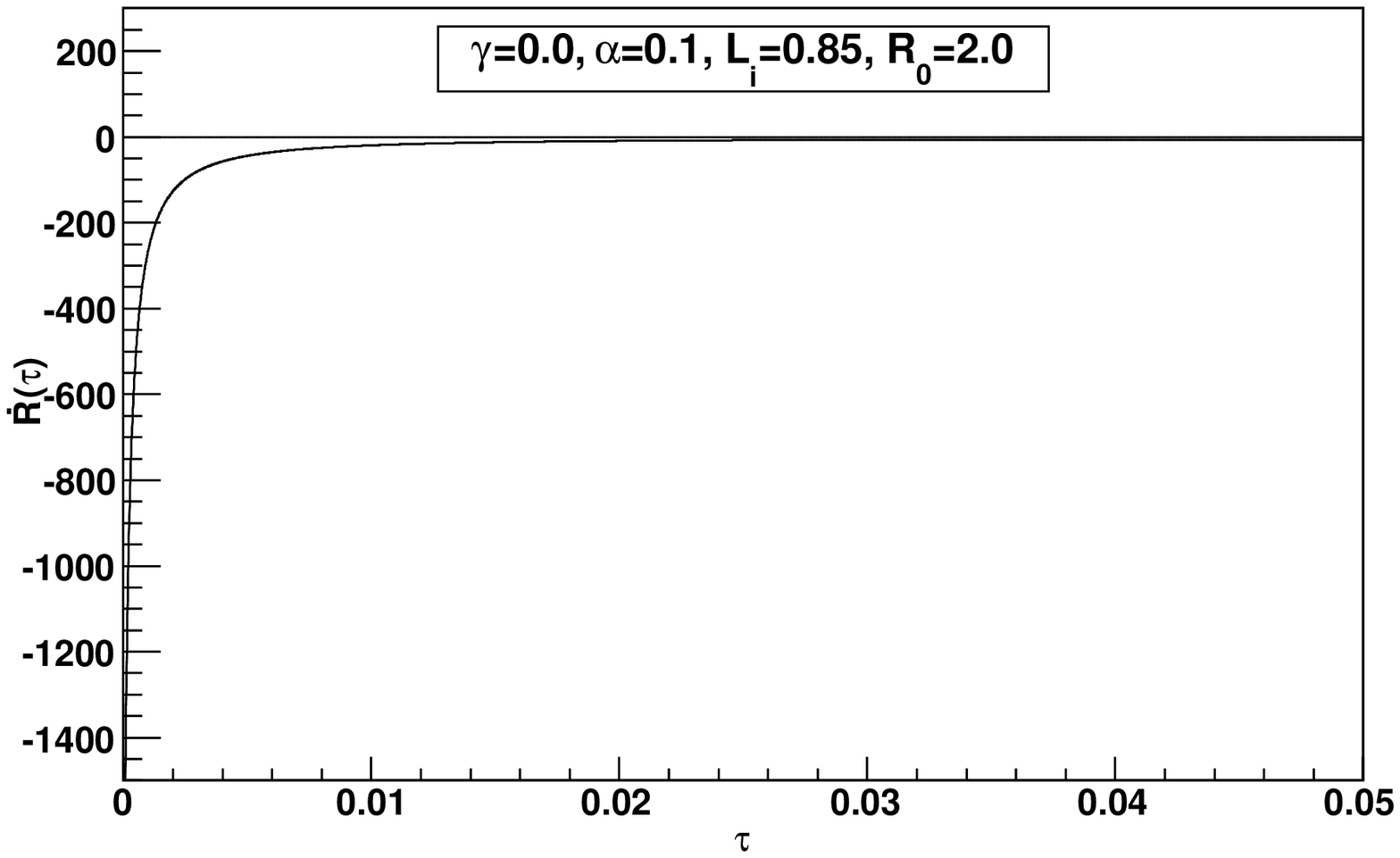}{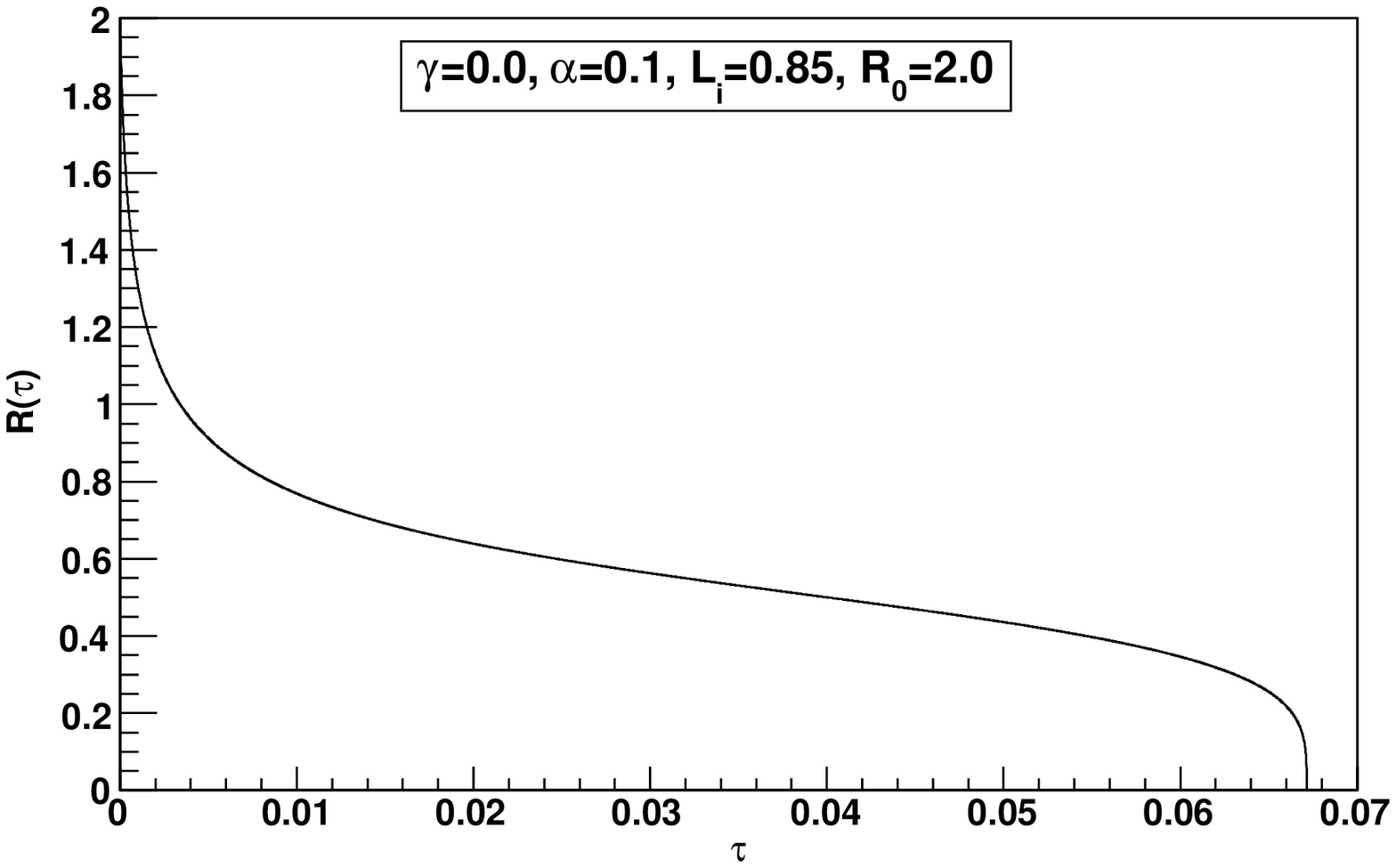}
\caption{These figures show the dynamical evolution of collapse
of these zero mass gravastars, forming at the end of the evolution a point-like zero mass object.
We have assumed the values $\alpha=0.1$, $L_i=0.85$, $\gamma=0$ and initial radius $R_0=R(\tau=0)=2$.}
\lb{Rtau}
\end{figure}

\begin{figure}
\epsscale{.80}
\plotone{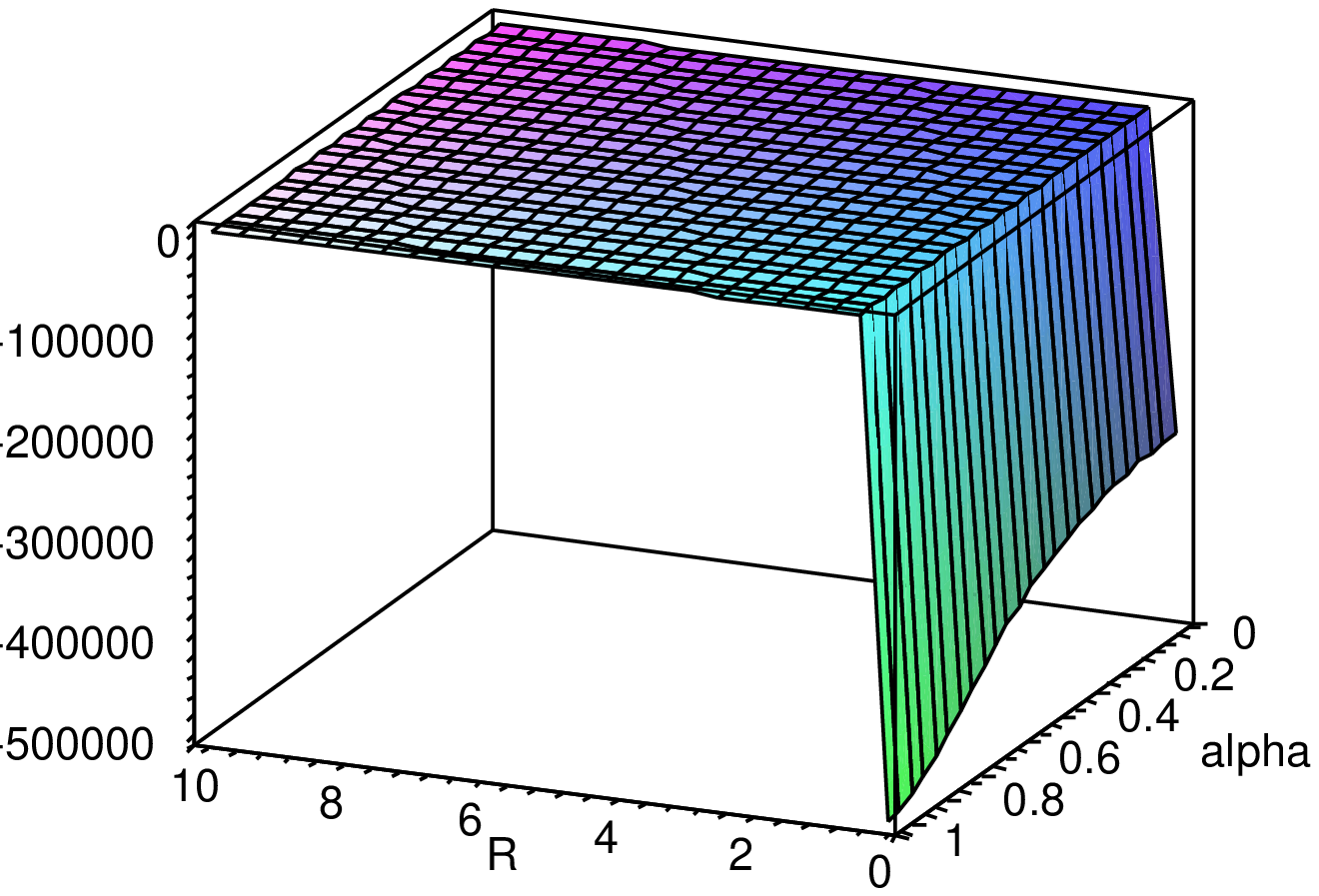}
\caption{This plot shows, in terms of $\alpha$ and $R$, the potential $V(R,L_i=L_{ic},\alpha,\gamma=0)$, in the intervals $0 < R < 10$ and $0 < \alpha < 1$, because $R< L_{ic}$. Since the potential is always negative anywhere, this shows that the potential $V(R,L_i,\alpha,\gamma=0)$ represents a collapse.}
\lb{fig6}
\end{figure}

\begin{figure}
\epsscale{.80}
\plotone{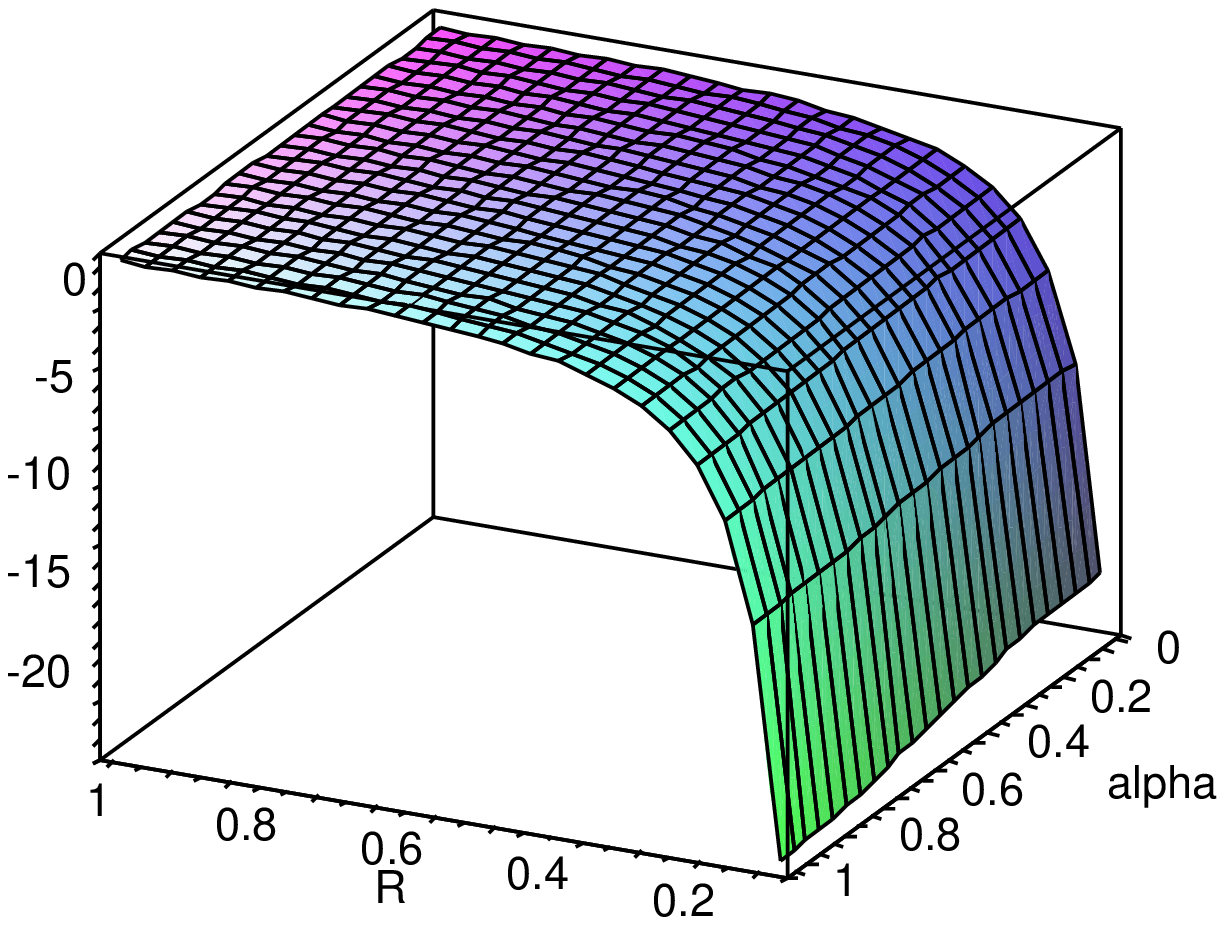}
\caption{This plot shows, in terms of $\alpha$ and $R$,
the potential $V(R,L_i=L_{ic},\alpha,\gamma=1)$, in the intervals
$0 < R < 1$ and $0 < \alpha < 1$, because $R < L_{ic}$. Since the potential is always negative
anywhere, this shows that the potential $V(R,L_i,\alpha,\gamma=1)$ represents a
collapse.}
\lb{fig7}
\end{figure}

\begin{figure}
\epsscale{.80}
\plotone{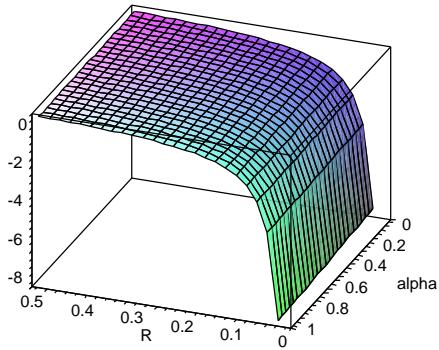}
\caption{This plot shows, in terms of $\alpha$ and $R$,
the potential $V(R,L_i=L_{ic},\alpha,\gamma=5/4)$, in the intervals
$0 < R < 0.5$ and $0 < \alpha < 1$, because $R < L_{ic}$. Since the potential is always negative
anywhere, this shows that the potential $V(R,L_i,\alpha,\gamma=5/4)$ represents a
collapse.}
\lb{fig8}
\end{figure}

Thus, the unique possible solution from gravastars models with $m=0$
represents a collapsing configuration which can produce a massive, but not
gravitationally interactive object, with an exterior de Sitter spacetime.

\section{Final Remarks}

In this paper we interpret a particular and very interesting solution,
which appears from the gravastars models. This corresponds to the case where
the Schwarzschild mass, which characterizes the exterior vacuum spacetime,
is zero. Then, we have a massive thin  shell with a de Sitter spacetime
inside, which generates a vacuum de Sitter exterior spacetime. 
Moreover, we can conclude that this structure is always unstable and
it collapses, although it does not form a black hole.
Instead of it reduces itself to a point-like object. This can be seen directly  from the analysis 
of the potential and confirmed by the relation between the cosmological 
constants ($L_i > L_e$), which is imposed in order to have a positive shell 
mass, and which violates the condition which had been already proposed by two 
of us, in another paper \citep{JCAP3}, that is, $L_i < L_e$ to have stable structures. It is remarkable that 
if $\Lambda_i = \Lambda_e$ there is no shell (its mass is zero), although it is not true if $m\neq 0$ .

This object does not interact gravitationally with an outside test particle. Then
it is similar to a point-like topological defect, since topological defects are
not only stable against small perturbations, but they cannot decay or be undone 
or be detangled, precisely because there is no continuous transformation that will map them 
(homotopically) to a uniform or "trivial" solution. 

\begin{acknowledgments}
We thank the helpful discussions with Dr. Anzhong Wang and to Dr. Pedro Rocha
for making some of the plots.
The financial assistance from FAPERJ/UERJ (MFAdaS) are gratefully acknowledged.
The author (RC and MFAdaS) acknowledges the financial support from FAPERJ (no. 
${\rm E}-26/171.754/2000$,
${\rm E}-26/171.533/2002$, ${\rm E}-26/170.951/2006$, ${\rm E}-26/110.432/2009$ and
${\rm E}-26/111.714/2010$). The authors (RC,
MFAdaS and JFVdR) also acknowledge the financial support from Conselho Nacional de Desenvolvimento Cient\'ifico e
Tecnol\'ogico - CNPq - Brazil (no. 450572/2009-9, 301973/2009-1 and 477268/2010-2). The author (MFAdaS)
also acknowledges the financial support from Financiadora de Estudos e Projetos - FINEP - Brazil
(Ref. 2399/03).
\end{acknowledgments}

\end{document}